\documentclass[aps,prb,twocolumn,amssymb,color,superscriptaddress,pdflatex]{revtex4-1}
\usepackage{color,physics}%needed for 'bcomment'
\usepackage{epstopdf,xcolor,graphicx,hyperref,physics,amsmath,comment}

\usepackage{amsmath}

\newcommand{\boldk}{\boldsymbol{k}}
\newcommand{\dsum}{\displaystyle \sum}
\newcommand{\disp}{\displaystyle}

\begin{document}
\title{Modeling multiorbital effects in Sr$_{2}$IrO$_{4}$ under strain and a Zeeman field}
\author{Lena Engstr\"om}
\affiliation{Department of Physics and the Centre for the Physics of Materials, McGill University, Montr\'eal, Qu\'ebec,
	H3A 2T8, Canada}
\affiliation{D\'epartement de Physique, Universit\'e de Montr\'eal, Montr\'eal, Qu\'ebec, H3C 3J7, Canada}
\affiliation{Regroupement Qu\'eb\'ecois sur les Mat\'eriaux de Pointe (RQMP)}
\author{T.~Pereg-Barnea}
\affiliation{Department of Physics and the Centre for the Physics of Materials, McGill University, Montr\'eal, Qu\'ebec,
	H3A 2T8, Canada}
	\affiliation{Regroupement Qu\'eb\'ecois sur les Mat\'eriaux de Pointe (RQMP)}
\author{William Witczak-Krempa}
\affiliation{D\'epartement de Physique, Universit\'e de Montr\'eal, Montr\'eal, Qu\'ebec, H3C 3J7, Canada}
\affiliation{Regroupement Qu\'eb\'ecois sur les Mat\'eriaux de Pointe (RQMP)}
\affiliation{Centre de Recherches Math\'ematiques, Universit\'e de Montr\'eal; P.O. Box 6128,
Centre-ville Station; Montr\'eal, Qu\'ebec, H3C 3J7, Canada}

\date{\today}
\begin{abstract}
We present a comprehensive study of a three-orbital lattice model suitable for the layered iridate Sr$_2$IrO$_4$. Our analysis includes various on-site interactions (including Hubbard and Hund's) as well as compressive strain, and a Zeeman magnetic field. We use a self-consistent mean field approach with multiple order parameters to characterize the resulting phases. While in some parameter regimes the compound is well described by an effective $J=1/2$ model, in other regimes the full multiorbital description is needed. As a function of the compressive strain, we uncover two quantum phase transitions: first a continuous metal-insulator transition, and subsequently a first order magnetic melting of the antiferromagnetic order. Crucially, bands of both $J=1/2$ and $J=3/2$ nature play important roles in these transitions. Our results qualitatively agree with experiments of Sr$_2$IrO$_4$ under strain induced by a substrate, and motivate the study of higher strains.
\end{abstract}
\maketitle
\section{Introduction}

The combination of strong correlations, spin-orbit coupling (SOC), and multiple relevant orbitals has proven to lead to many interesting states including spin- and orbital- orders, topological states and unconventional superconductivity \cite{Witczak-Krempa2014, Sato2015, Kargarian2011, Georges2013, Rau2016, Chen2020}. The iridate family of compounds displays a very rich phenomenology due to a combination of all of these factors\cite{Witczak-Krempa2014, Cao2018, Bertinshaw2019}. The five $d$-orbitals are usually split by crystal fields into two groups, $e_g$ and $t_{2g}$, with two-fold and three-fold degeneracy respectively. On the other hand, strong spin-orbit coupling may lead to further energy splitting which in turn may reduce the number of relevant bands. Early works on the iridates noted that the spin-orbit coupling affects the system to such an extent that the local total angular momentum states, referred to here as $J$-eigenstates, do not mix. Moreover, the strong SOC allows one to project onto the $J=1/2$ subspace and arrive at a simplified effective one-orbital model. In this work we go beyond this effective $J_{\text{eff}}=1/2$ model and examine regimes where considering a larger subspace, with multiple orbitals, is deemed necessary. 

\begin{figure} 
\centering 
\includegraphics[width=8.6cm]{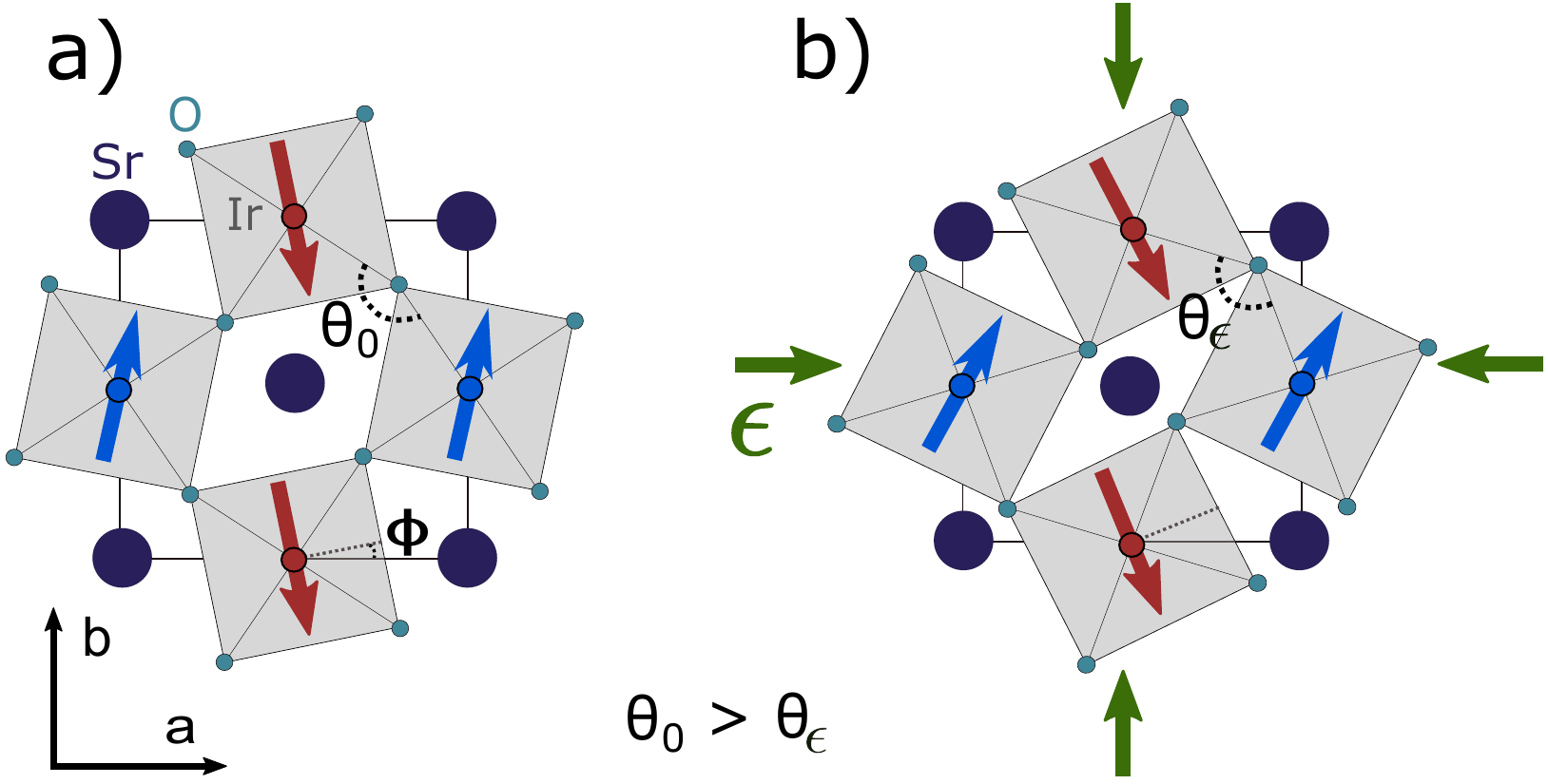}
\caption{a) The structure of a single layer of Sr$_{2}$IrO$_{4}$ without strain, $\epsilon =0$. The IrO$_{6}$ octahedra are rotated in-plane by an angle of $\phi \approx \pm 12^{\circ}$ with the sign opposite on neighboring octahedra. This yields an angle $\theta_{0} \approx 156^{\circ}$. The arrows represent the total magnetic moments in the ground state, $\boldsymbol{\mu} = - \mu_{\text{B}}\left( \boldsymbol{L} + g \boldsymbol{S} \right)$, which are arranged in a canted antiferromagnetic fashion with a small net moment along the $a$-axis. b) When compressive strain, $\epsilon$, is applied to the layer, the angle $\theta_{\epsilon}$ decreases as the rigid octahedra are rotated closer together. A tensile strain has the opposite effect, resulting in a larger angle $\theta_{\epsilon}$. }
\label{fig:structure}
\end{figure}

Sr$_{2}$IrO$_{4}$ is the single-layer compound in the Ruddlesden-Popper series of perovskite iridates and is a spin-orbit coupled Mott insulator with a canted antiferromagnetic order, as seen in Fig.~\ref{fig:structure}. In each layer the iridium atoms are arranged in a square lattice. Each iridium site is surrounded by an oxygen octahedron which is rotated with respect to the crystallographic axes, by a staggered angle $\phi \approx \pm 12^{\circ}$\cite{Boseggia2013}. The magnetic moment roughly follows the rotation of each octahedron, resulting in the canted order. In this state the system's properties are dominated by the $J=1/2$ bands, which are separated from the $J= 3/2$ bands\cite{Kim2008,Jackeli2009, Wang2011}. A projected effective model therefore seems appropriate. This view is further supported by the x-ray absorption spectra that indicate scattering paths corresponding to an order formed by $J = 1/2$ pseudospins\cite{Kim2009}.

The appropriate effective one-orbital model is surprisingly similar to the one used successfully to describe many of the features of the cuprate high-T$_c$ superconductors. A three-orbital model can take into account both the $J=1/2$ and $J=3/2$ subspaces. Previous studies of this multiorbital model of Sr$_{2}$IrO$_{4}$ predict that superconductivity could occur in this compound as well. However, d-wave superconductivity seems only possible for interorbital interaction parameters in the lower end of the predicted range\cite{Meng2014, Nishiguchi2019, Yang2014}. These predictions indicate that the effective one-orbital model, $J_{\text{eff}}=1/2$, might only be valid in some regimes. The system enters other regimes when effects, such as of doping, are no longer small compared to the energy scale of the spin-orbit coupling.

In this paper we take the approach that the three-orbital model is necessary. Including the six bands of the three $t_{2g}$ orbitals, allows us to study several regimes where the effective one-orbital model may be insufficient. We consider the effects of an epitaxial strain and an external magnetic field on {\it undoped} Sr$_{2}$IrO$_{4}$. Strain and a Zeeman field are both orbital dependent effects: the strain deforms the lattice and changes the inter-orbital overlaps; the Zeeman field couples to the magnetic moment which depends on the orbital as well as the spin angular momentum. 

When considering strain, we should note that Sr$_{2}$IrO$_{4}$ is sensitive to changes in lattice geometry via a strong Jahn-Teller effect\cite{Liu2019}. Epitaxial strain affects the lattice constants as well as the rotation angle $\phi$. Strain is introduced by growing Sr$_{2}$IrO$_{4}$ on a substrate with a mismatch in lattice parameters\cite{Lupascu2014, Hao2019, Miao2014, Geprags2020}. In Sr$_{2}$IrO$_{4}$, an epitaxial strain which changes the lattice parameters by $0.5\%$ is not only easily achievable but also enough to reduce the N\'eel temperature by $30$K\cite{Lupascu2014, Hao2019}. Epitaxial strain is thus a suitable handle for tuning interactions and lattice deformations. \textit{Ab initio} calculations have previously identified contributions from different $J$-states to the experimentally observed magnetic order, as well as excitations between the states for some strain values\cite{Bhandari2019, Seo2019}. Compressive epitaxial strain mainly modifies the lattice structure by increasing the rotation angle of the octahedra surrounding the iridium sites, see Fig.~\ref{fig:structure}b.

The same effect can be achieved by other means. Two recent promising methods to modify the rotation angle, are electrical current\cite{Cao2020a} and ``field altering'' via growth in a magnetic field\cite{Cao2020b}. In particular, the method of ``field altering'' in combination with doping has recently been proposed to provide a more favorable environment for observing superconductivity in Sr$_{2}$IrO$_{4}$\cite{Cao2020b}. These experiments motivate us to study trends for a range of strain values and a range of interaction parameters.

Another regime where it might be important to include all three orbitals is reached when a Zeeman field is applied. The field couples to the total magnetic moment which is a combination of the orbital and spin angular momentum, and therefore mixes the local $J$-states. This mixing has been largely neglected in previous literature as the Zeeman field effects were studied in the context of the effective $J=1/2$ model\cite{Wang2011, Carter2012, Takayama2016, Seifert2019, Porras2019}. Previously, both experiments and modelling of the Sr$_{2}$IrO$_{4}$ compound have observed a metamagnetic transition at small fields\cite{Kim2009, Cao1998, Nauman2017, Rathi2018, Rodan2018}. This transition aligns the canting of the antiferromagnetic order between layers in the compound, at a field around 0.3T \cite{Liu2019, Haskel2012}. In this work we consider higher fields as we expect to be able to see effects originating from in-plane interaction within each layer after the metamagnetic transition has taken place.

Some recent work with orbital resolved measurements in a magnetic field has, in addition, shown unequal contributions from each of the $t_{2g}$ orbitals to the magnetic moment\cite{Jeong2019}. For the simpler $J_{\text{eff}}=1/2$ projected model, contributions from each orbital are assumed to be equal. This motivates our choice to study the three-orbital model in a Zeeman field.

In this work we aim to give further insight into how quantum phase transitions can arise in Sr$_{2}$IrO$_{4}$ under a compressive epitaxial strain, with the addition of a Zeeman field. In section \hyperref[sec:Model]{II} we introduce a three-orbital Hubbard-Kanamori model with on-site interactions. The interactions are treated with a self-consistent mean field approximation. The mean field decoupling includes all possible uniform and staggered order parameters, except superconductivity. We include a Zeeman field which is applied in different directions and couples to the full magnetic moment $\boldsymbol{\mu} = - \mu_{\text{B}}\left( \boldsymbol{L} + g \boldsymbol{S} \right)$, where $\mu_{\text{B}}$ is the Bohr magneton. The compressive epitaxial strain is modelled as a linear change in hopping parameters. This allows us to reach higher compressive epitaxial strain than previously modeled. We are considering a 2-atom unit cell in the canted lattice, as in Fig.~\ref{fig:structure}, where the mean field order parameters are calculated without assuming any relation between the two sublattices. A set of 42 independent order parameters is therefore used. These parameters describe order in the orbital and spin angular momentum and can be expressed in the $J$-state basis or the orbital basis. By considering the full set of order parameters the contributions to the order from each $J$-state as well as contributions from order parameters mixing $J$-states, are considered. Section \hyperref[sec:Results]{III} presents the results where our model predicts phase transitions from an insulating antiferromagnet into metallic states at high strains. In section \hyperref[sec:StrainDrivenPT]{III.A} details are given for the transitions which are induced by a compressive strain. The Fermi surfaces for the metallic orders are predicted to include several $J$-states, highlighting the necessity of the multiorbital model. In section \hyperref[sec:OrbitalContributions]{III.B} the contributions to the magnetic moment from our set of order parameters are considered when a field is applied. Changes to the contributions of order parameters from different $J$-states are predicted as a function of strain and field. Finally, in section \hyperref[sec:Discussion]{IV} we relate our results to experimental findings and discuss implications of entering regimes where the $J_{\text{eff}}=1/2$ model is insufficient.

\section{Model} \label{sec:Model}
In Sr$_{2}$IrO$_{4}$, the octahedral crystal fields around the iridium splits its $d$-levels into $t_{2g}$ and $e_g$ orbitals. Without doping, the three $t_{2g}$ orbitals, $d_{yz}$, $d_{xz}$, and $d_{xy}$, are filled with five electrons while the $e_g$ orbitals are unoccupied at higher energy. Besides the intra- and inter-orbital hopping, these atomic states are also subject to a large on-site spin-orbit coupling and interactions. While the Hubbard interaction strength $U$ is rather moderate, around $1- 2$eV, the spin-orbit coupling (SOC) is strong, $\lambda \approx 0.4$eV. The strong SOC splits the six $t_{2g}$ bands roughly in two groups: four bands of mainly $J=3/2$ character and two bands of mainly $J=1/2$ character. In the undoped compound the Fermi level is placed in such a way that the $J=3/2$ bands are filled and $J=1/2$ bands are half-filled. The interaction strength is therefore enough to form an AFM state dominated by the $J=1/2$ pseudospins\cite{Kim2009}. This state is depicted in Fig.~\ref{fig:structure}. The anisotropy of the system causes the interactions to be significantly stronger in the plane than out-of-plane. A combination of the anisotropy and the in-plane staggered rotations of the iridium sites causes the magnetic order to form in the plane along the crystallographic $b$-axis with a canting angle of the magnetic moment along the $a$-axis in each plane. In this work, given the large anisotropy, we model the system as a single layer.

\subsection{Hubbard-Kanamori Model} \label{sec:HKmodel}
Before we introduce the strain and Zeeman field, we recall the Hamiltonian of the system:
\begin{equation}
H= H_{\text{kin}} + H_{\text{SOC}} + H_{\text{I}}
\end{equation}
where $H_{\text{kin}}$ is the kinetic part, $H_{\text{SOC}}$ is the spin-orbit coupling, and $H_{\text{I}}$ contains the on-site interactions, as defined below. The kinetic part includes hopping between nearest and next nearest neighbouring sites for each of the d-orbitals $\alpha= yz,xz,xy $, with inter- and intra-orbital hopping. In order to study uniform and staggered orders we consider a unit cell with two sites, with sublattices $s=A,B$. The sublattices include the staggered rotation $\phi_{s}= \pm \phi$, with opposite signs for sublattice $A$ and $B$. For both sublattices defined in the same global basis $\boldsymbol{c}=$ $( c_{A, yz, \uparrow},$ $ c_{A, yz, \downarrow},$ $c_{A, xz, \uparrow},$ $c_{A, xz, \downarrow},$ $c_{A, xy, \uparrow},$ $c_{A, xy, \downarrow},$ $c_{B, yz, \uparrow},$ $c_{B, yz, \downarrow},$ $c_{B, xz, \uparrow},$ $c_{B, xz, \downarrow},$ $c_{B, xy, \uparrow},$ $c_{B, xy, \downarrow})$, the labelling of orbital and spin directions are along the crystallographic $a$- and $b$-axes. The rotation of each site can be taken into account in the kinetic Hamiltonian which therefore includes non-zero hoppings between the $d_{yz}$ and $d_{xz}$ orbitals. Our Hamiltonian follows the form of Ref.~[\onlinecite{Carter2013}], which uses a Slater-Koster approach\cite{Slater1954}. For each spin $\sigma=\uparrow, \downarrow$ the kinetic terms take the form (in momentum space):
\begin{equation}
H_{\text{kin}} = \left( \begin{array}{c c }
H_{AA} & H_{AB} \\
H_{AB}^{\dagger} & H_{BB}
\end{array} \right)
\end{equation}
\begin{equation}
H_{AA} = \left( \begin{array}{c c c}
\epsilon_{d} & \epsilon_{1d} & 0 \\
\epsilon_{1d} & \epsilon_{d} & 0 \\
0 & 0 & \epsilon^{xy}_{d} 
\end{array} \right), H_{AB} = \left( \begin{array}{c c c}
 \epsilon_{yz} & - \epsilon_{rot}&0 \\
 \epsilon_{rot}& \epsilon_{xz} & 0 \\
 0 & 0 & \epsilon_{xy}
\end{array} \right)
\end{equation}
where
\begin{equation}
\begin{aligned}
\epsilon_{xy}&= 2 t \left( \cos k_{x} + \cos k_{y} \right)\\
 \epsilon_{yz}&= 2 \left( t_{\delta} \cos k_{x} + t_{1} \cos k_{y} \right) \\
\epsilon_{xz}&= 2 \left( t_{1} \cos k_{x} + t_{\delta} \cos k_{y} \right)\\ 
 \epsilon_{rot}&= 2 t' \left( \cos k_{x} + \cos k_{y} \right)\\
\epsilon^{xy}_{d}& = 4 t_{n} \cos k_{x} \cos k_{y} +\mu_{xy}\\
\epsilon_{1d}&= 4 t_{1d} \sin k_{x} \sin k_{y} \\
\epsilon_{d}&= 4 t_{nd} \cos k_{x} \cos k_{y}.
\end{aligned}
\end{equation}
The nearest-neighbor hopping for $d_{yz}$- and $d_{xz}$-orbitals is nearly one dimensional in-plane, with $t_{1}$ along the direction in which they are orientated and a smaller $t_{\delta}$ along the other direction. The $d_{yz}$-$d_{xz}$ inter-orbital hopping, $t'$, and the nearest-neighbor hopping between $d_{xy}$-orbitals, $t$, are equal in both directions. For the next-nearest-neighbors, along the diagonal of the square lattice, the hopping is $t_{n}$ for $d_{xy}$ and $t_{nd}$ for the $d_{yz}$- and $d_{xz}$-orbitals. The $d_{yz}$-$d_{xz}$ inter-orbital hopping is $t_{1d}$ along the diagonal. In the absence of strain we use the following values: $(t, t_{1}, t_{\delta}, t', t_{n}, t_{1d}, t_{nd})=$ \\ (-0.211, -0.186, -0.055, -0.042, -0.118, -0.004, 0.021)eV. These values are extrapolated from those calculated for compressive epitaxial strain by the lineraziation given in detail below in section~\ref{sec:EpitaxialStrain}. The hopping amplitudes have been calculated by Seo \textit{et al}.\cite{Seo2019} through \textit{ab initio} for varying strain. The corresponding rotation angle of the sites is $\phi_{s}=\pm 12.3^{\circ}$ and $\mu_{xy}=0.7t$\cite{Mohapatra2020, Bertinshaw2019} takes the tetragonal splitting into account, with the value of $t$ being fixed to that of $\epsilon=0$. In general, the tetragonal splitting is expected to change under compression as the tetragonal elongation of the oxygen octahedra increases\cite{Haskel2012}. Works considering a superexchange Hamiltonian predict that for an increased elongation, either an order along the $c$-axis can be favoured or the canting moment can be suppressed\cite{Perkins2014, Liu2015}. An additional small staggering of the distortion has been observed to stabilize the canted magnetic moment\cite{Torchinsky2015}. However, we chose to study the strain-induced hopping modifications separately as there are conflicting predictions on how the energy splitting depends on strain. \textit{Ab initio} calculations predicted a $\mu_{xy}$ where the absolute value decreases until $\mu_{xy}$ changes sign \cite{Zhang2013, Bhandari2019}, while recent RIXS data observed a linearly increasing absolute value of $\mu_{xy}$ \cite{Paris2020}. Section~\ref{sec:Discussion} expands on how strain-dependent distortions could affect our results.

The atomic spin-orbit interaction, with the coupling $\lambda$, is defined at each site from spin and orbital angular momentum along the same axes as:
\begin{equation}
\begin{array}{r l}
H_{\text{SOC}}=& \disp \frac{\lambda}{2} \dsum_{\boldsymbol{j}, i} \dsum_{\alpha \beta, \sigma \sigma'} L^{i}_{\alpha \beta} \sigma^{i}_{\sigma \sigma'} c^{\dagger}_{ \boldsymbol{j} \alpha \sigma} c_{\boldsymbol{j} \beta \sigma'}
\end{array}
\end{equation}
where $i=x,y,z$, $\boldsymbol{\sigma}=  \left( \sigma^{x},\sigma^{y},\sigma^{z} \right)$ are the Pauli matrices in the spin basis $\sigma=\uparrow, \downarrow$, and the matrices
\begin{equation}
\boldsymbol{L} = \left( \begin{bmatrix}
 0 & 0 & 0 \\
 0 & 0 & -i \\
 0 & i& 0
\end{bmatrix}, \begin{bmatrix}
 0 & 0 & i \\
 0 & 0 & 0 \\
 -i & 0& 0
\end{bmatrix},\begin{bmatrix}
 0 & -i & 0 \\
 i & 0 & 0 \\
 0 & 0& 0
\end{bmatrix} \right)
\end{equation}
are the orbital angular momentum operators, projected onto the $t_{2g}$ subspace and written in the orbital basis $\alpha= yz, xz, xy $. The interactions in the multiband model on each site take the form of the Kanamori-Hubbard interactions\cite{Kanamori1963} 
\begin{equation}
\begin{array}{r l}
H_{\text{I}}= &\displaystyle U \dsum_{\boldsymbol{j}, \alpha} n_{\boldsymbol{j} \alpha \uparrow} n_{\boldsymbol{j} \alpha \downarrow} \\ &
+ \displaystyle \dsum_{\boldsymbol{j}, \alpha \neq \beta} J_{\text{H}} \left[ c^{\dagger}_{\boldsymbol{j} \alpha \uparrow} c^{\dagger}_{\boldsymbol{j} \beta \downarrow} c_{\boldsymbol{j} \alpha \downarrow} c_{\boldsymbol{j} \beta \uparrow} + c^{\dagger}_{\boldsymbol{j} \alpha \uparrow} c^{\dagger}_{\boldsymbol{j} \alpha \downarrow} c_{\boldsymbol{j} \beta \downarrow} c_{\boldsymbol{j} \beta \uparrow} \right] \\ &
+ \displaystyle \dsum_{\boldsymbol{j}, \alpha < \beta, \sigma} \left[ U' n_{\boldsymbol{j} \alpha \sigma} n_{\boldsymbol{j} \beta \bar{\sigma}} + \left(U' - J_{\text{H}} \right) n_{\boldsymbol{j} \alpha \sigma} n_{\boldsymbol{j} \beta \sigma} \right]
\end{array}
\end{equation}
with the intraorbital interactions $U$, the Hund's coupling $J_{\text{H}}$, and the interorbital repulsion $U'$. For simplicity the spherically symmetric value $U' = U- 2 J_{\text{H}}$ is taken. For Sr$_{2}$IrO$_{4}$ the Hund's coupling is approximated to be in the range $0.05U-0.2U$ \cite{Meng2014, Nishiguchi2019, Yang2014, Zhou2017}.

\subsection{Zeeman Coupling} \label{sec:ZeemanCoupling}
We consider the effect of an external magnetic field $\boldsymbol{H}$ through the Zeeman field. The field couples to the full magnetic moment $\boldsymbol{\mu} = \mu_B\left(\boldsymbol{L}+g\boldsymbol{S}\right) $, with $g\approx 2$ being the gyromagnetic ratio. The additional term in the Hamiltonian is
\begin{equation}\label{eq:Zeeman}
\begin{array}{r l l}
H_{\text{Z}} = & \displaystyle \mu_{\text{B}}\sum_{ \boldsymbol{j}, s} \sum_{\alpha, \sigma} &\left[ \dsum_{\beta} \boldsymbol{H} \cdot \boldsymbol{L}_{\alpha \beta} c^{\dagger}_{s, \boldsymbol{j} \alpha \sigma } c_{s, \boldsymbol{j} \beta \sigma} \right. \\ & & \left. \disp + \frac{1}{2} \sum_{\sigma'}g \boldsymbol{H} \cdot \boldsymbol{\sigma}_{\sigma \sigma'} c^{\dagger}_{s,\boldsymbol{j} \alpha \sigma} c_{s,\boldsymbol{j} \alpha \sigma'} \right].
\end{array}
\end{equation}
For realistic magnetic fields, the Zeeman energy is significantly smaller than the spin-orbit coupling $\lambda \approx 0.4$eV, and the gap $ \approx 0.5$eV. For example, a field of $H \approx 10$T corresponds to an energy of the order of $g \mu_{\text{B}} H =1.2$meV.
\begin{figure*} 
\centering 
\includegraphics[width=17.2cm]{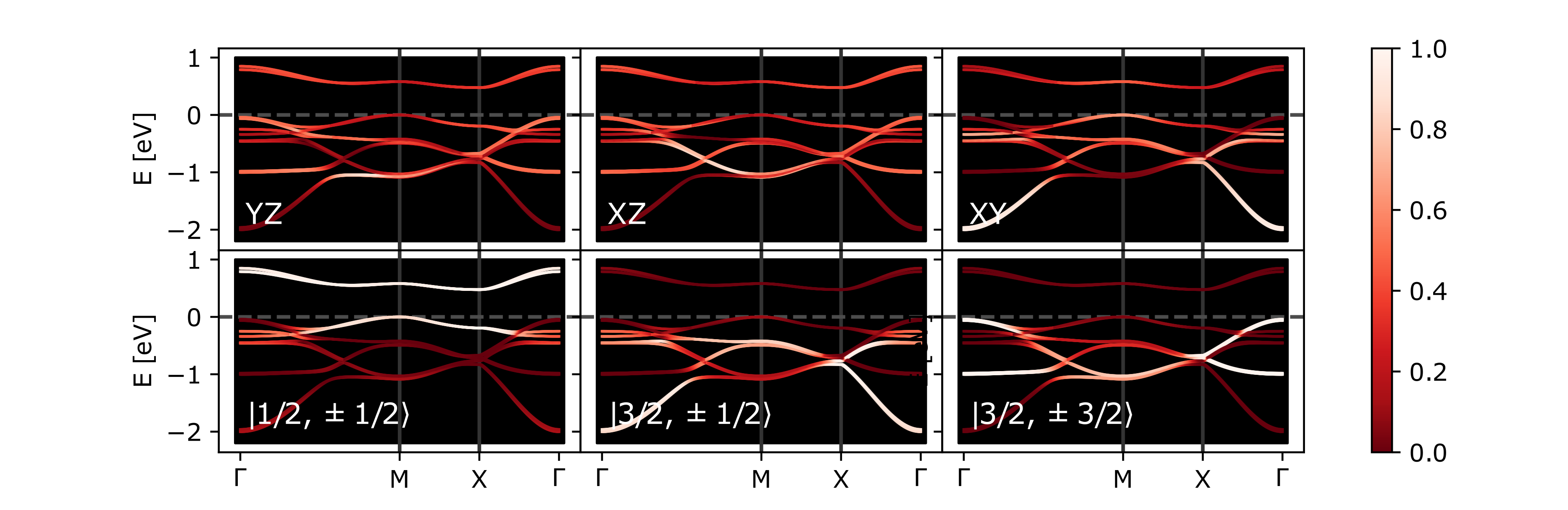} 
\caption{The band structure is shown for the antiferromagnetic insulating state in Sr$_{2}$IrO$_{4}$, calculated for $\lambda=0.38$eV, $U=0.9$eV, and $J_{\text{H}}/U=0.1$, with no applied strain or field. As there are 6 states per site, the band structure consists of 12 bands forming a staggered order. In the top row, the weight from each orbital $d_{yz}$, $d_{xz}$, and $d_{xy}$ is projected onto the eigenstates at each $k$-point in the Brillouin zone as in Eq.~\eqref{eqn:Porb}. The large spin-orbit coupling mixes the orbitals, so the bands closest to the Fermi level have contributions from all three orbitals. The second row shows the eigenstates projected onto the $J=1/2$ and $J=3/2$ states as in Eq.~\eqref{eqn:Pj}. The $J=1/2$ bands dominate near the Fermi level except near $\Gamma$, where $J=3/2$ takes over.}
\label{fig:bandStrucs}
\end{figure*}

\subsection{Epitaxial Strain} \label{sec:EpitaxialStrain}
We model the effect of a compressive strain on the system by modifying the hopping parameters linearly with the strain. We use a linearization of the set of values for the hopping parameters calculated by Seo \textit{et al}.\cite{Seo2019}. In Ref.~[\onlinecite{Seo2019}], the compound is grown on three different substrates which have lattice constants that are smaller than that of Sr$_{2}$IrO$_{4}$: (LaAlO$_{3}$)$_{0.3}$(Sr$_{2}$TaAlO$_{6}$)$_{0.7}$, NdGaO$_{3}$, and LaAlO$_{3}$. The resulting misfit strain modifies the lattice constants in the Sr$_{2}$IrO$_{4}$ thin film. X-ray diffraction measurements find these modified lengths and \textit{ab initio} calculations are performed for those structures. The calculations therefore provide three data points for the hopping parameters at given values of the compressive strain. In this work we use those three data points to fit a linear dependence of the hopping with the strain. Our linearization results in the proportional changes, $\rho$, which modify our hopping amplitudes as
\begin{equation} \label{eqn:strainT}
\begin{array}{c}
t (\epsilon)= t \left(1 + \rho \epsilon \right)\\
t_{1} (\epsilon)= t_{1} \left(1 + \rho_{1} \epsilon \right) \\
t' (\epsilon)= t' \left(1 + \rho' \epsilon \right) \\
t_{n} (\epsilon)= t_{n} \left(1 + \rho_{n} \epsilon \right)\\
t_{\delta} (\epsilon)= t_{\delta} \left(1 + \rho_{\delta} \epsilon \right)\\
t_{1d} (\epsilon)= t_{1d} \left(1 + \rho_{1d} \epsilon \right)\\
t_{nd} (\epsilon)= t_{nd} \left(1 + \rho_{nd} \epsilon \right)\\
\phi(\epsilon)= \phi \left(1 + \rho_{\phi} \epsilon \right).
\end{array}
\end{equation}
For a compressive strain ($\epsilon <0 $) the resulting values are $(\rho,\rho_{1}, \rho', \rho_{n}, \rho_{\delta}, \rho_{1d}, \rho_{nd}, \rho_{\phi} ) =$ (0.014,-0.251, -0.309, -0.048, 0, 0,-0.02,-0.085). The values used for $\epsilon=0$ are those given by this linearisation. As illustrated in Fig.~\ref{fig:structure}, the effect of compressive strain is mainly to increase the relative rotation angle between adjacent octahedra. However, by using these values we are not restricted to consider only rotation effects. The rotations change the overlap integrals between orbitals on different sites. The nearest neighbor inter-orbital $d_{yz}$-$d_{xz}$ hopping, as well as the next nearest neighbor intra-orbital $d_{xy}$ hopping are increased under strain. On the other hand, the nearest neighbor $d_{xy}$ hopping is decreased. Our linearized strain model allows us to predict what orders can arise when we reach strain values beyond the experimentally achieved $\epsilon=-1.9\%$\cite{Seo2019}.

\subsection{Mean Field Approximation} \label{sec:MFApprox}
In mean field theory one approximates the Hamiltonian by a quadratic one, so that the quartic interaction terms are decomposed by introducing a variety of order parameters. This yields an auxiliary Hamiltonian for which the spectrum can be found by diagonalizing a single-particle Hamiltonian. The resulting eigenstates are then used as variational states to calculate the expectation value of the original {\it interacting} Hamiltonian for a given electron density. The energy is minimized with respect to the order parameters, thus determining their values. With two atoms per unit cell, three orbitals and two spin states, each unit cell has 12 creation/annihilation operators. A mean field order parameter is the expectation value of a bilinear operator $\langle c^{\dagger}_{\alpha} c_{\beta} \rangle$. Our mean field decomposition is done by choosing to include the full set of on-site order parameters under the condition of a hermitian auxiliary/mean-field Hamiltonian. For each of the sites in the unit cell we form a $6 \times 6$ hermitian matrix of order parameters, meaning that we calculate a total of $2 \cdot 21 = 42$ independent complex-valued order parameters. The set of order parameters is therefore $\langle c^{\dagger}_{\gamma_{1}} c_{\gamma_{2}} \rangle _{s}$, where $\gamma_{i}$ is a label combining the spin label $\sigma$ and the orbital label $\alpha$ in each sublattice $s=A,B$. The order parameters are calculated in iterative steps through the coupled set of self-consistency equations, as given in Appendix \hyperref[sec:appA]{A}. The calculated order parameters are used as input into the Hamiltonian in order to repeat the process in iterative steps until the input and output, of the form presented in table \ref{table:OPs}, differ by less than the total tolerance of $10^{-5}$. The calculations were performed on a $200 \times 200$ grid of momentum $k$-points. A range of initial conditions are considered to ensure that the global minimum of the energy functional is found.

Our analysis assumes no relations between the order parameters on the different sites. Uniform orders are considered by calculating the net value of the order parameters from both sites, $(A+B)/2$, and staggered orders are the difference in order parameters between sites, $(A-B)/2$. Such staggered orders include commensurate charge density waves (CDW), spin density waves (SDW), orbital density waves (ODW), and spin-orbit density waves (SODW). It is convenient to rewrite the order parameters in order to directly describe the spin and orbital angular momentum. The order parameters $n_{yz}$, $n_{xz}$, and $n_{xy}$ are the filling of each orbital. The spin $S^{i}$ and the orbital angular momentum $L^{i}$ are calculated in each direction $i=x,y,z$. Order parameters that couple spin and orbital degrees of freedoms, like the bare SOC, $\Lambda^{i}$ are included as well. Suppressing the sublattice label, these order parameters are given by:
\begin{equation}\label{eq:SpinOrbitOPs1}
S^{i}_{\alpha} = \displaystyle \frac{1}{2} \sum_{\sigma, \sigma'} \sigma_{ \sigma \sigma'}^{i} \langle c^{\dagger}_{\alpha \sigma} c_{\alpha \sigma'} \rangle
\end{equation}
\begin{equation}\label{eq:SpinOrbitOPs2}
L^{i}_{\sigma} = \displaystyle \sum_{\alpha, \beta} L_{ \alpha \beta}^{i} \langle c^{\dagger}_{\alpha \sigma} c_{\beta \sigma} \rangle
\end{equation}
\begin{equation}\label{eq:SpinOrbitOPs3}
\Lambda^{i} = \displaystyle \frac{1}{2} \sum_{\alpha, \beta} \sum_{\sigma, \sigma'} L_{ \alpha \beta}^{i} \sigma_{ \sigma \sigma'}^{i} \langle c^{\dagger}_{\alpha \sigma} c_{\beta \sigma'} \rangle.
\end{equation}

Once a set of self-consistent order parameters has been found in the orbital and spin basis, they can also be expressed in the $J$-basis. This basis represents the eigenstates of the non-interacting model in the $\lambda \rightarrow \infty$ limit, in which the hopping can be neglected. Order parameters expressed in this basis represent contributions of each $J$-state as well as a measure of the mixing between states. The transformation $\tilde{c}_{ m, \tau} = \sum_{\alpha, \sigma} U^{\alpha, \sigma}_{ m, \tau} c_{\alpha, \sigma}$, generates the basis $\tilde{c}_{m, \tau}$ at each site where $m=| j, j^{z} \rangle:$ $1=|1/2,\pm 1/2 \rangle, 2=|3/2, \pm 1/2 \rangle, 3=|3/2, \pm 3/2 \rangle$ are the pseudospins and $\tau=+,-$. The same transformation is applied for both sublattices, which defines the $J$-states in the global basis. It is important to note that $J$-states that are defined for \textit{local} rotated orbitals are different states and such a definition may slightly shift the resulting contributions of each state. In the $J$-basis, order parameters are constructed as a linear combination of the ones discussed above in Eqs.~\eqref{eq:SpinOrbitOPs1},\eqref{eq:SpinOrbitOPs2},\eqref{eq:SpinOrbitOPs3}. These order parameters are given by $\langle \tilde{c}^{\dagger}_{m, \tau} \tilde{c}_{n, \tau'}\rangle$ and are transformed from the orbital basis as:
\begin{equation} \label{eqn:Utransf}
\langle \tilde{c}^{\dagger}_{m \tau} \tilde{c}_{n \tau '} \rangle = \dsum _{\alpha, \beta, \sigma, \sigma'} \left( U^{\alpha \sigma}_{m \tau} \right) ^{\ast} U^{\beta \sigma '}_{n \tau '} \langle c^{\dagger}_{\alpha \sigma} c_{\beta \sigma '} \rangle
\end{equation}
with the matrix $U$ given in Appendix \hyperref[sec:appB]{B}. In this basis we consider the order parameters:
\begin{equation} \label{eqn:Jm}
J^{i}_{m} = \displaystyle \frac{1}{2} \sum_{\tau, \tau'} \sigma_{ \tau \tau'}^{i} \langle \tilde{c}^{\dagger}_{m \tau} \tilde{c}_{m \tau'} \rangle
\end{equation}
\begin{equation} \label{eqn:Jmn}
J^{i}_{mn} = \displaystyle \frac{1}{2} \sum_{\tau, \tau'} \sigma_{ \tau \tau'}^{i} \langle \tilde{c}^{\dagger}_{m \tau} \tilde{c}_{n \tau'} \rangle
\end{equation}
for the $J$-states $m,n=1,2,3$, and the pseudospins $\tau, \tau' = +,-$. In addition, the filing of each $J$-state is given by
\begin{equation} \label{eqn:nmOPs}
n_{m} = \displaystyle \sum_{\tau} \langle \tilde{c}^{\dagger}_{m \tau} \tilde{c}_{m \tau} \rangle.
\end{equation}
This transformation extends the analysis of Mohapatra and Singh in Ref.~[\onlinecite{Mohapatra2020}], who studied the contributions $J_{m}$, without strain and a Zeeman field. In this work we include the additional mixing $J_{mn}$, which includes effects beyond those that can be projected onto the individual subspaces of the $J$-states. The amount of mixing $J_{mn}$ allows us to see whether strain and Zeeman fields require us to go beyond the effective $J_{\text{eff}}=1/2$ model.
%%%%%%%%%%%%%%%%%%%%% 

\section{Results} \label{sec:Results}
\begin{table*}[]
\centering
\begin{tabular}{|l| cccccccccc|}
\hline 
 & $n_{yz}$ & $n_{xz}$ & $n_{xy}$ & $L$ & $S_{yz}$ & $S_{xz}$ & $S_{xy}$ & $\Lambda_{x}$ & $\Lambda_{y}$ & $\Lambda_{z}$ \\
 \hline
staggered & 0 & 0 & 0 & -0.47 & 0.13 & -0.15 & 0.11 & 0 & 0 & 0\\
net & 1.66 & 1.63 & 1.71 & 0.15 & 0.040 & -0.042 & -0.037 & 0.32 & 0.30 & 0.35\\
\hline
 & $n_{1}$ & $n_{2}$ & $n_{3}$ & $J_{1}$ & $J_{2}$ & $J_{3}$ & $J_{12}$ & $J_{13}$ & $J_{23}$ & \\
 \hline
staggered & 0 & 0 & 0 & -0.29 & -0.0047 & 0.0040 & -0.023 & 0.011 & 0.0005 & \\
net & 1.02 & 1.99 & 1.99 & 0.12 & 0.0018 & 0.0012 & 0.0035 & -0.0010 & -0.0002 & \\
\hline
\end{tabular}
\caption{ The order parameters are given in the three-orbital basis, as in Eqs.~\eqref{eq:SpinOrbitOPs1}, \eqref{eq:SpinOrbitOPs2}, \eqref{eq:SpinOrbitOPs3}, as well as in the basis of $J$-states, as in Eqs.~\eqref{eqn:Jm}, \eqref{eqn:Jmn}. The calculation is performed at $\lambda=0.38$eV, $U=0.9$eV, and $J_{\text{H}}/U=0.1$, with no strain or field, meaning that the state is the canted antiferromagnet in Fig.~\ref{fig:structure}a. The differences between order parameters in the two sublattices are given as the staggered value. The net values of the order parameters are defined as the average for the two-site unit cell. For the $L$ and $S_{yz,xz,xy}$ order parameters, the staggered values are along the $b$-axis, while the net values are along the $a$-axis. The order parameters $\Lambda$ renormalize the spin-orbit coupling strength.}
\label{table:OPs}
\end{table*}

First, our mean field solution in the absence of Zeeman field and strain is in agreement with previous studies\cite{Mohapatra2020, Meng2014, Nishiguchi2019, Yang2014, Zhou2017, Kim2008, Watanabe2010,Wang2011}. In Fig.~\ref{fig:bandStrucs} we present the band structure for $\lambda=0.38$eV, $U=0.9$eV, and $J_{\text{H}}/U=0.1$. Under these conditions, both this work and other studies, find a band gap close to the experimentally observed value\cite{Witczak-Krempa2014}. The resulting state is an antiferromagnet along the $b$-axis with a small staggered canting angle of $\phi_{\mu}\approx \pm 14^{\circ}$ along the $a$-axis. This angle is larger than the rotation of the underlying lattice and slightly larger than what is observed in experiments\cite{Boseggia2013}. The magnetic order canting angle does not precisely match the lattice rotation angle due to the tetragonal distortion and a non-zero Hund's coupling. An angle difference is captured by our model and even by the projected $J=1/2$ model\cite{Jackeli2009}. The resulting eigenstates are expressed in the two bases, the orbital and the $J$-basis, and the contributions of each state can be calculated at all $k$-points for each band. For orbitals defined in the \textit{global} basis the eigenstates $|n (\boldk) \rangle$ can be expressed in the components $|n (\boldk) \rangle = \dsum_{\alpha, \sigma, s} \eta_{\alpha, \sigma,s, n} (\boldk) |\alpha, \sigma, s \rangle$. The transformation onto the $J$-basis is done for each site individually in the global basis with the matrix $U$ given in \eqref{eqn:Umat} in Appendix \hyperref[sec:appB]{B}, $|n (\boldk) \rangle = \dsum_{m, \tau, s} \eta'_{m, \tau,s, n} (\boldk) |m, \tau, s \rangle =\dsum_{m, \tau, s} \dsum_{\alpha, \sigma} \eta'_{m, \tau,s, n} (\boldk) \left( U^{\alpha \sigma}_{m \tau} \right)^{\ast} |\alpha, \sigma, s \rangle$.
The weight of an orbital in an eigenstate at a given $k$-point is calculated as
\begin{equation}\label{eqn:Porb}
P_{n,\alpha}(\boldsymbol{k}) =\displaystyle\sum_{s=A,B} \sum_{\sigma=\uparrow, \downarrow}|\eta_{ \alpha, \sigma, s, n} (\boldsymbol{k})|^{2},
\end{equation}
in the original three-orbital basis and:
\begin{equation}\label{eqn:Pj}
P_{n,m}(\boldsymbol{k}) =\displaystyle\sum_{s=A,B} \sum_{\tau=+,-}|\eta'_{ m, \tau, s, n} (\boldsymbol{k})|^{2}, 
\end{equation}
in the $J$-state basis.
The values are displayed for the full bandstructure in Fig.~\ref{fig:bandStrucs} and the figure is complemented by the values of the order parameters in Table \ref{table:OPs}. The magnetic order receives the largest contribution from the $J=1/2$ states, as given by Eq.~\eqref{eqn:Jm}. Similarly, as can be seen in the lower panels of Fig.~\ref{fig:bandStrucs}, the $J=1/2$ states are dominant in the two bands closest to the Fermi level, except near the $\Gamma$-point. Expressed in the orbital basis, the same bands are a mixture of all three orbitals, with the contribution of $d_{xy}$ being slightly smaller. Additional bands that appear close to the Fermi level, at the $\Gamma$-point, are bands of $|3/2, \pm 3/2 \rangle $ character. However, Table \ref{table:OPs} shows that these states offer only a small contribution to the AFM order. Similarly, the order parameters which mix the $|1/2, \pm 1/2 \rangle $ and the $|3/2, \pm 1/2 \rangle$ states have a contribution of about 5-10$\%$ of the one of $J=1/2$, which is not negligible. A similar discrepancy in the magnetic order has been identified previously\cite{Kim2008} by observing a larger ratio of orbital angular momentum, compared to spin angular momentum, than expected from a pure $J=1/2$ order.

\subsection{Strain-Driven Phase Transitions} \label{sec:StrainDrivenPT}
\begin{figure}
\centering 
\includegraphics[width=8.6cm]{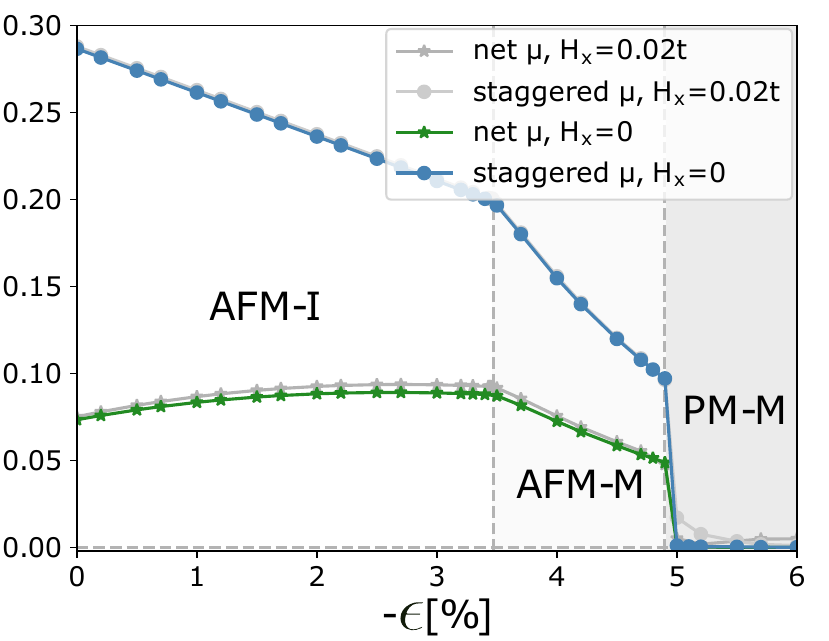} 
\caption{The total magnetic moment is plotted for increasing compressive strain. The staggered moment is the difference between the two sublattices and the net moment results from the canting of the moments at both sites. The strain-driven transitions take place both under zero field as well as under a large Zeeman field $H_{x}=0.02t$, along the $a$-direction. The critical strain values, at which the gap closes and the system first goes into a metallic AFM (AFM-M) and subsequently into a paramagnetic (PM-M) state, are marked for zero field by vertical dashed lines. As shown in Fig.~\ref{fig:PDsepH}, these transitions are only slightly shifted by the field. When a field is applied there is a small remaining AFM moment, below $2 \cdot 10^{-2}$, that appears right after the transition into the PM-M state. The evolution of the bandgap with strain is shown in Fig.~\ref{fig:gapStrain}.}
\label{fig:MomentumHx}
\end{figure}

In this subsection we discuss the effects of strain. The magnetic moment for both the staggered AFM order and the net moment is shown in Fig.~\ref{fig:MomentumHx}. As the compressive strain is increased the antiferromagnetic order decreases and two phase transitions occur. At lower strain values the staggered magnetic moment in the insulating (AFM-I) order continuously decreases until the gap closes, in a continuous Lifshitz transition into an antiferromagnetic metal (AFM-M). The strain dependence of the band gap is plotted in Fig.~\ref{fig:gapStrain} in Appendix \hyperref[sec:appC]{C}. As the strain increases further, the antiferromagnetic order continues to decrease until a strain value where a first order transition into a paramagnetic metal (PM-M) occurs. The transitions are driven by the increasing bandwidth of the $J=1/2$ bands and an increase in the energy of the $J=3/2$ bands. We will describe several multiorbital aspects of the strain-driven phase transitions: (i) the changes in multiorbital contributions close to critical strain, (ii) the additional bands contributing to the Fermi surface in the metallic state, and (iii) the dependence of the critical strain on model parameters.

\begin{figure} 
\centering 
\includegraphics[width=8.6cm]{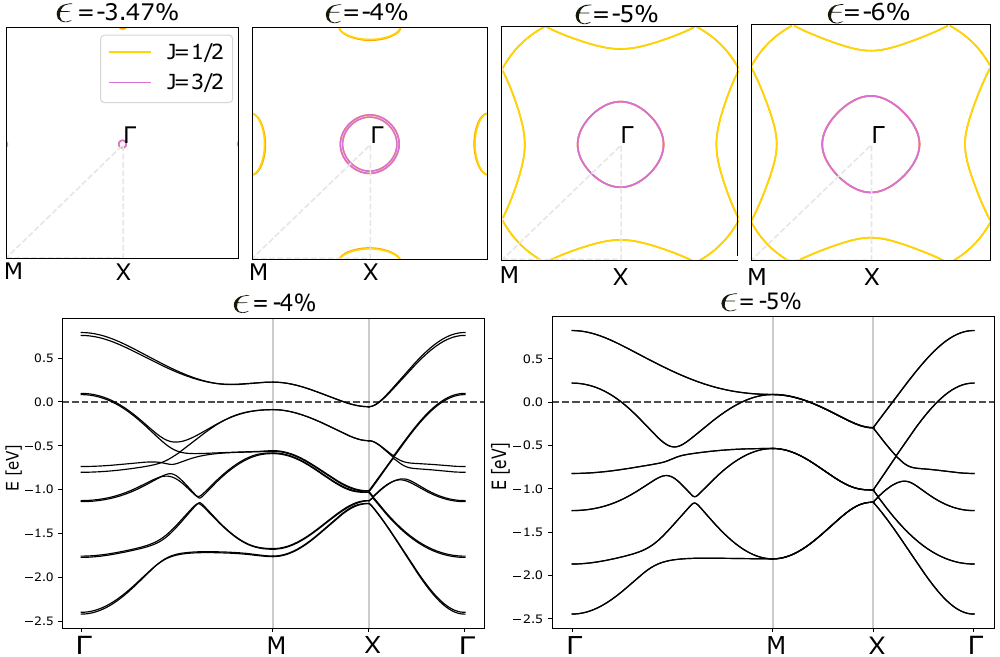} 
\caption{Fermi surfaces $(E=0)$ for the strain-driven transitions at zero field. The dominant bands are identified as being of mainly $| j, j^{z} \rangle = | 1/2, \pm 1/2 \rangle $ and of $| j, j^{z} \rangle = | 3/2, \pm 3/2 \rangle $ character, by the same method as in Fig \ref{fig:FSprojection}. As the strain is increased the indirect gap in the AFM order decreases and eventually closes at $\epsilon=-3.47 \% $. For Fermi surfaces in the metallic AFM (AFM-M) phase, such as at $\epsilon=-4 \%$, some band splitting can be observed. The splitting occurs at these points as the resulting FM component corresponds to a larger canting angle than the underlying rotation of the lattice. At $\epsilon=-5 \% $, the system becomes a paramagnetic metal and an additional $J=1/2$ surface appears around the $M$-point. }
\label{fig:FSstrain}
\end{figure}

\begin{figure} 
\centering 
\includegraphics[width=8.6cm]{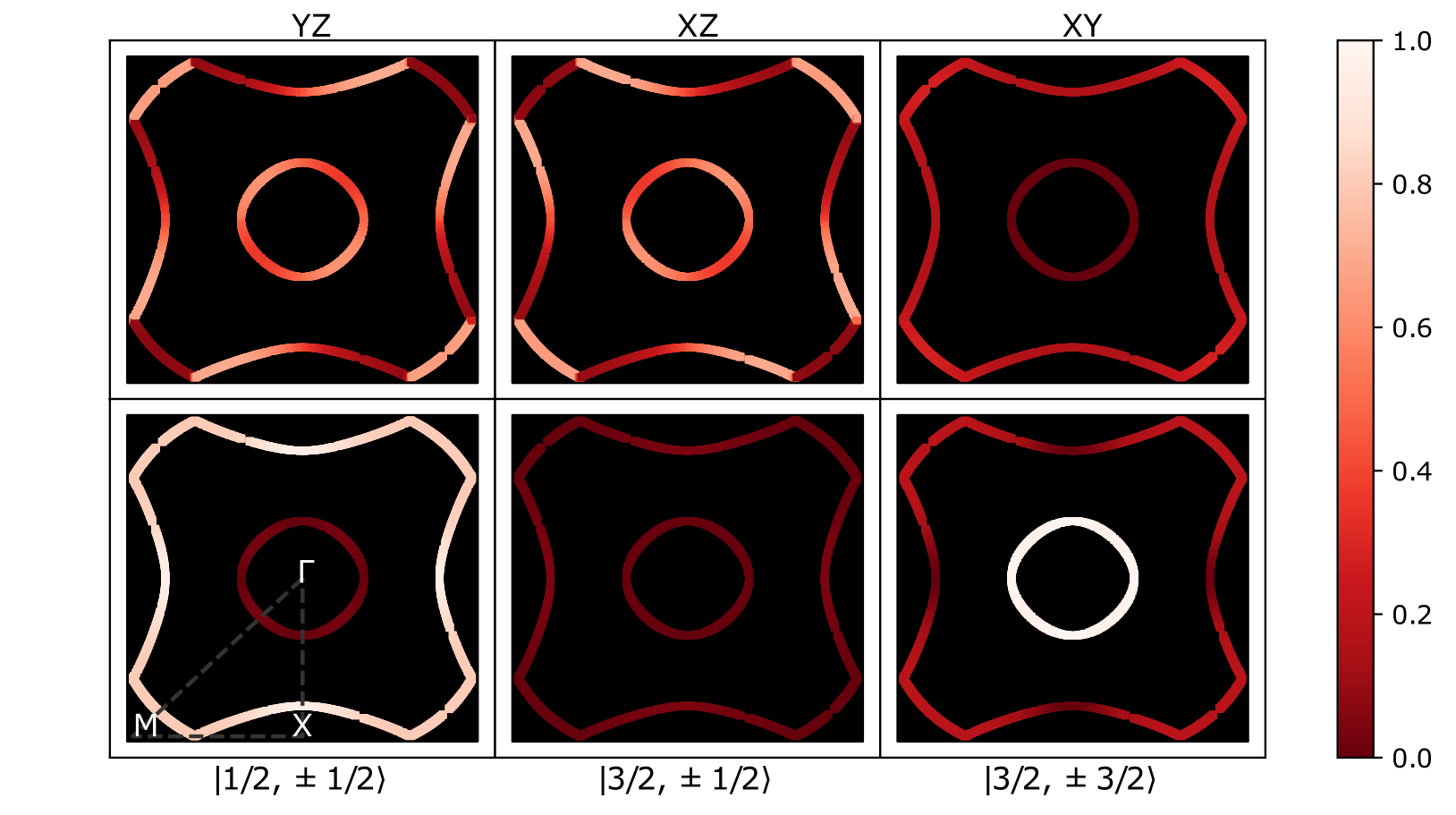}
\caption{The Fermi surface at zero field and a compressive strain $\epsilon= -5\%$ is shown with the calculated contributions from each orbital, in the upper row, and from each $J$-state, in the lower row. As in Fig.~\ref{fig:bandStrucs}, orbital weights for each state are calculated for eigenstates at each $k$-point in the Brillouin zone, according to Eqs.~\eqref{eqn:Porb}, \eqref{eqn:Pj}. As shown in Fig.~\ref{fig:FSstrain} the bands can be described mainly by the $| j, j^{z} \rangle = | 1/2, \pm 1/2 \rangle $ states around the $M$- and $X$-points, and by the $| j, j^{z} \rangle = | 3/2, \pm 3/2 \rangle $ states around the $\Gamma$-point.}
\label{fig:FSprojection}
\end{figure}

Approaching the first transition by increasing the strain, we see a decrease in the staggered magnetic moment. The decrease is mostly felt in the $J=1/2$ subspace, and therefore the relative contribution of the $J=3/2$ states to the magnetic order is increased. As the underlying rotations of the lattice increase, so does the canting angle of the antiferromagnetic state. The changes in orbital contributions are discussed further in \hyperref[sec:OrbitalContributions]{III.B}. At higher strains in the metallic state, several bands cross the Fermi level. The resulting Fermi surfaces are shown in Fig.~\ref{fig:FSstrain} for several strain values. Different parts of the Fermi surface have a different character, as shown in Fig.~\ref{fig:FSprojection}. In this figure both possible bases are projected onto the Brillouin zone. Pockets around the $M$- and $X$-points are clearly dominated by the $J=1/2$ states. However, another pocket near the $\Gamma$-point originates from a band with a high $|3/2, \pm 3/2 \rangle $ contribution. In the orbital basis, the pockets can be described as alternating sections of $d_{yz}$ and $d_{xz}$ orbitals, where the sections dominated by each orbital are related by a rotation of $\pi/2$, see Fig.~\ref{fig:FSprojection}.

The two phase transitions, as indicated in Fig.~\ref{fig:MomentumHx}, are determined to occur at $\epsilon = -3.47\%$, the point at which the indirect gap closes, and at $\epsilon =-4.9\%$, where the order parameters for the staggered magnetic moment become lower than $2 \cdot 10^{-2}$. The Fermi surfaces appearing at lower strain values have small pockets of $J=1/2$ and $J=3/2$ character which gradually increase in size as the strain increases. In the AFM phase, the canting angle of the AFM order is larger than the rotation angle of the underlying lattice. As a result, a small band splitting can be observed close to the $\Gamma$-point for the pockets of $J=3/2$ character. As the size of the pocket increases at higher strain values and the AFM order decreases, this splitting is decreased. In the paramagnetic phase an additional pocket of $J=1/2$ character appears at the $M$-point.

The value of the critical compressive strain that we obtain as the transition point between metallic and insulating magnetically ordered states depends on our model parameters. Fig.~\ref{fig:gapStrain} in Appendix \hyperref[sec:appC]{C} shows a range of critical compressive strains for other possible values of the interaction $U$. In our model, the critical strain value mainly depends on the size of the initial gap. Therefore, the critical strain increases with spin-orbit coupling and with the interaction $U$, and decreases with the Hund's coupling $J_{\text{H}}$. The agreement between experimental work and our predictions for the decreasing AFM order as a function of strain, as well as possible values for a realistic critical strain are discussed below. 

When a Zeeman field is applied only minimal changes to the critical strain are observed. This is shown in the phase diagrams in Fig.~\ref{fig:PDsepH} in Appendix \hyperref[sec:appD]{D}. Additional effects to orbital contribution from a magnetic field are discussed in the following section.

\subsection{Orbital contributions} \label{sec:OrbitalContributions}
\begin{figure}
\centering 
\includegraphics[width=4.2cm]{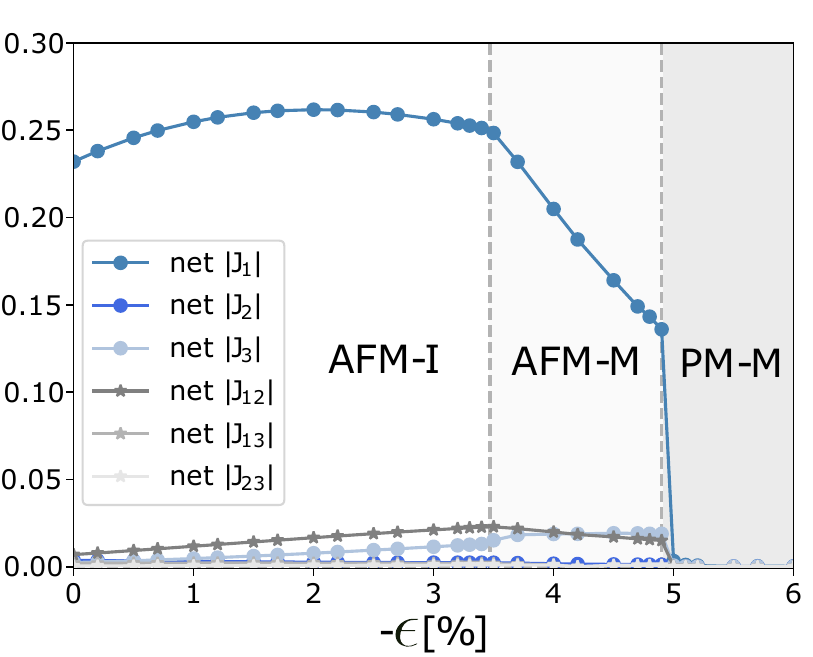} 
\includegraphics[width=4.2cm]{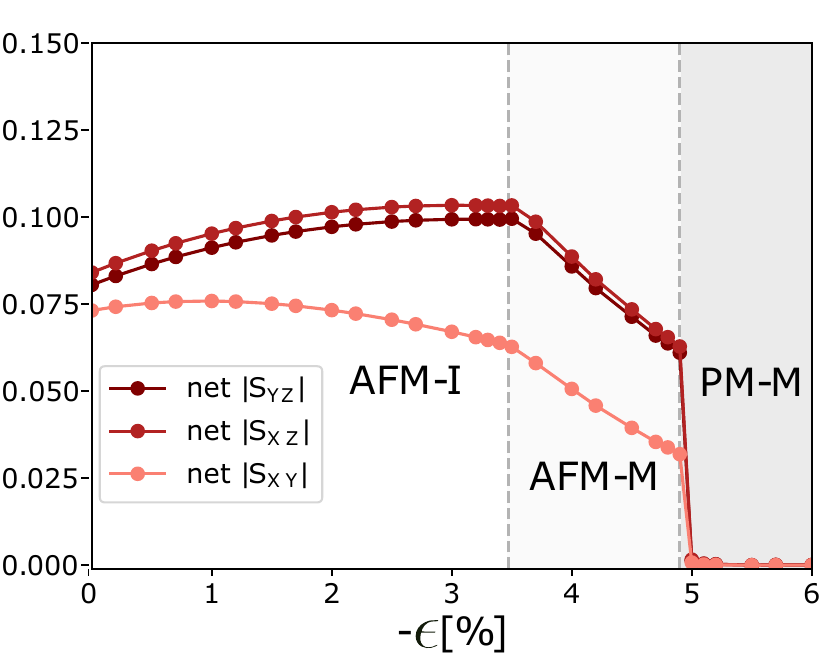}
\includegraphics[width=4.2cm]{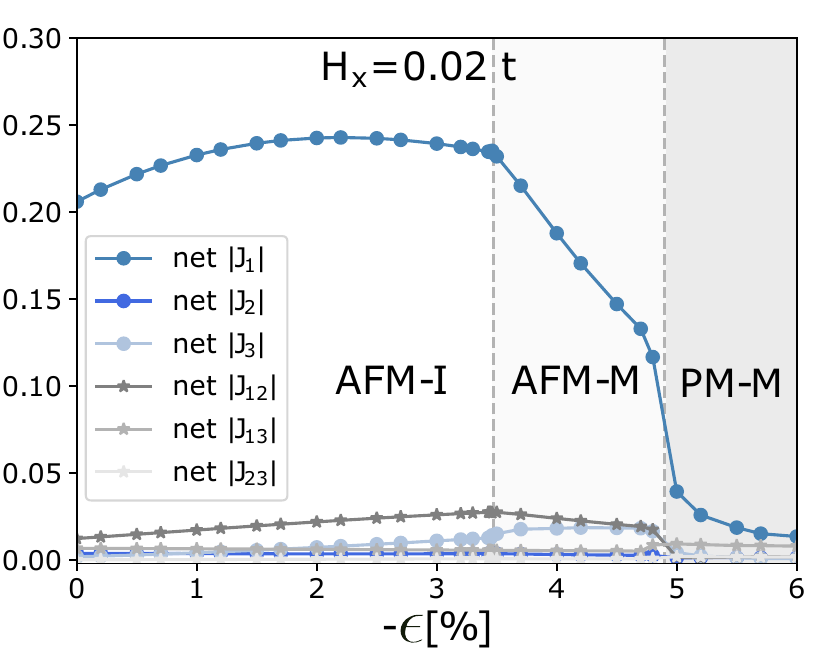} 
\includegraphics[width=4.2cm]{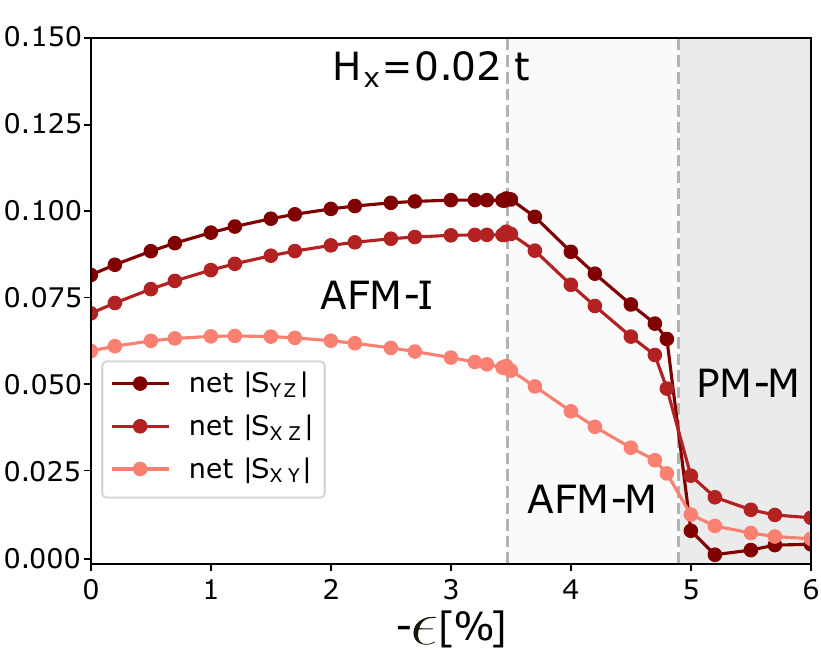}
\caption{Order parameters for the net magnetization are plotted for an increasing compressive strain, both in the $J$-basis and as spin contributions from each orbital. These plots display the strain-driven transitions into metallic states shown in Fig.~\ref{fig:MomentumHx}. In the $J$-basis, order parameters $J_{m}$ for each state and order parameters $J_{mn}$ mixing $J$-states, as in Eqs.~\eqref{eqn:Jm} and \eqref{eqn:Jmn}, are shown. There are minor changes in the contributions from each order parameter with strain, before the transition out of the insulating antiferromagnetic (AFM-I) state. However, once a field is applied there is a clear difference in contributions to the net moment between AFM and PM orders. }
\label{fig:contributions}
\end{figure}

At the strain-driven phase transitions depicted in Fig.~\ref{fig:MomentumHx}, contributions from the $J$-states, $J_{m}$, and the mixing between those states, $J_{mn}$, change by different amounts. The contributions from the spin angular momentum and the $J$-states to the net moment are shown in Fig.~\ref{fig:contributions}, both without an applied field and for a Zeeman field in-plane along the $a$-axis ($H_{x}$). The figure shows how the strain and the Zeeman field affect the magnetic order. As the insulating AFM order decreases under strain, the order in $J=1/2$ decreases while the order in other states remain roughly constant. While strain increases the staggered rotation angle of the AFM state and therefore all $J$-states, the Zeeman field tends to affect orbitals depending on their relative orientation to the field.

The changes in contributions to the net moment under strain are minor. The net moment increases as the staggered AFM order follows the increased underlying staggered rotation of the octahedra surrounding the Ir sites. In the metallic AFM order the contribution from the $J=1/2$ states to the net moment mainly decreases while the others remain constant. When a high in-plane field is applied there are additional distinguishing effects between the AFM and the PM. In the insulating AFM state there is some increased mixing contributions to the net moment, as the field does not couple purely to the $J$-states. The $J=1/2$ states however still clearly dominate in the antiferromagnetic phase.

For the orbital angular momentum basis, the spin order $S_{\alpha}$, in each orbital, $\alpha$, is also plotted in Fig.~\ref{fig:contributions}. For zero field the orbitals start out with close to equal spin order and as the strain is increased the $S_{xy}$ order decreases. When the in-plane field is applied, the AFM-I state has a larger contribution from the $S_{yz}$ order while this dominance does not remain in the paramagnetic state. For an out-of-plane field $(H_{z})$ this results in a larger contribution from the $d_{xy}$-orbital, which corresponds to an increased mixing between $|1/2,\pm 1/2 \rangle$ and $|3/2,\pm 1/2 \rangle$ in the $J$-state basis. An in-plane field $(H_{x})$ increases contributions from the $d_{yz}$-orbital, or a mixing between the states $|1/2,\pm 1/2 \rangle$ and $|3/2,\pm 3/2 \rangle$.

In addition, in Fig.~\ref{fig:JnetlaJH} in Appendix \hyperref[sec:appC]{C} the parameters $\lambda$ and $J_{\text{H}}$ take on a range of possible values. At different values the amount of mixing between $J$-states (at zero strain) changes. The mixed $J$ order parameters, $J_{mn}$, in Eq.~\eqref{eqn:Jmn} are useful as they indicate whether a projected $J=1/2$ model is appropriate. Regimes with larger $J_{mn}$ values can therefore be identified as promising starting points for future studies of possible interband fluctuations and orders.

\section{Discussion} \label{sec:Discussion}
In this work, we have presented a mean field, zero temperature, analysis of the six-band Hubbard-Kanamori model for \textit{undoped} Sr$_{2}$IrO$_{4}$. A self-consistent mean field treatment considers a 2-atom unit cell and all 42 possible local order parameters. We study the undoped compound in the presence of both strain and a Zeeman field. In the absence of strain and field our model predicts an insulating canted antiferromagnetic state, in agreement with previous studies\cite{Mohapatra2020, Meng2014, Nishiguchi2019, Yang2014, Zhou2017, Kim2008, Watanabe2010,Wang2011} and experimental evidence\cite{Chikara2009, Fujiyama2012, Dai2014, Ye2013}. Upon applying a compressive strain our model predicts two transitions: a Lifshitz transition into an antiferromagnetic metallic state and, at higher strain, a first order transition into a metallic paramagnet. These transitions exist for a range of plausible interaction strengths. The inclusion of multiple bands is crucial to model these transitions. A decreased $J=1/2$ AFM order can in principle be described by projecting the effects of the strain onto the effective one-orbital $J_{\text{eff}}=1/2$ model. However, the strain causes the appearance of additional bands at the Fermi level that are missed by a $J_{\text{eff}}=1/2$ model.

Our predictions for the strain effects agree with trends from previous theoretical and experimental studies. For example, in Ref.~[\onlinecite{Hao2019}] the strain is shown to cause a decrease in the AFM order manifested in a lowered N\'eel temperature. As found in our model, the increased importance, due to strain, of the $J=3/2$ states also agrees with the observed intensity increase in optical transitions between $J=3/2$ and $1/2$ states found in other studies\cite{Zhang2013, Kim2016, Paris2020}. In addition, transport measurements observe a steady decrease in resistivity as the compressive epitaxial strain is increased\cite{Souri2019}. Such a trend can be expected from our calculations, as they predict a decreasing gap. At the highest measured strain value for epitaxial strain, $\epsilon=-1.9 \%$, the behavior is determined to still be insulating\cite{Souri2019}. Therefore, a transition has not been reached at that point. Our model predicts the same behavior. It is however important to note that generally mean field theory overestimates ordering. Fluctuations not taken into account here may shift the phase boundaries. Moreover, the interaction and spin-orbit coupling strength aren't directly measurable and we therefore choose parameters that match the previously found band structure\cite{Mohapatra2020, Zhou2017, Kim2008, Watanabe2010, Wang2011}. To get a range of possible strain values which will be relevant for future studies, a relation between possible initial gaps and the critical strain is given in Appendix \hyperref[sec:appC]{C}. 

Our results also include effects of various parameters on the mixing between different total angular momentum sectors. When the mixing between $J$-states is small, the $J_{\text{eff}}=1/2$ model can describe the ordered state well. However, for a larger mixing the full six-band model is necessary. We find that a larger strain, larger Hund's coupling, and lower spin-orbit coupling all increase the mixing. The Zeeman field also results in increased mixing, which depends on the direction of the field. It is worth noting that the mixing can be traced by studying the orbital content of each band. The orbital dependence of the magnetic state was recently determined, by Jeong {\it et al.} in Ref.~[\onlinecite{Jeong2019}], from the symmetry of occupied orbitals as measured by polarized neutron diffraction experiments. A similar experiment could potentially observe the strain-induced changes in orbital contributions found here.

The comparisons of our results to experiments with pressure are limited due to our one-layer model. For epitaxial strain/hydrostatic pressure, the distance between layers in the perovskite structure increases/decreases. Under pressure, the resulting increased interlayer interactions affect the magnetic order\cite{Haskel2020}. Additionally, our model may not be capturing all aspects of the strain-driven phase transitions. At high hydrostatic pressures, experiments are possibly pointing towards frustration from enhanced nearest- and next-nearest-neighbor interactions in an insulating quantum paramagnet\cite{Haskel2020}. Similarly as transport measurements not displaying any anomaly at the N\'eel temperature\cite{Cao2018}, studies considering hydrostatic pressure found a separation in the behaviour between magnetic order and insulating properties\cite{Haskel2012}, which is beyond the scope of our mean field theory. As can be seen in Fig.~\ref{fig:FSstrain}, we predict that several of the bands are located close to the Fermi surface during the strain-driven transitions. This regime could therefore potentially host strongly correlated interband effects.
 
The model considered in our work only describes compressive strain. There have however been several studies showing interesting effects at tensile strain or for other methods decreasing the rotation angle of the octahedra in Sr$_{2}$IrO$_{4}$, such as ``field altering'' or applying an electrical current\cite{Cao2020a, Cao2020b}. Experiments have shown both decreasing resistivity for tensile strain values\cite{Souri2019} and a lower N\'eel temperature for samples with a tensile strain of $\epsilon=0.4 \%$ than for those with a compressive strain of $\epsilon= -0.7 \%$ \cite{Seo2019}. However, \textit{ab initio} calculations at tensile strain\cite{Kim2016} pointed towards an increased charge gap which agrees with that observed in RIXS spectra\cite{Paris2020}. Accurately modelling the tensile regime might require the inclusion of additional effects. In future work, the strain value for which the pocket at the $\Gamma$-point appears in the Fermi surface could be adjusted by studying how the tetragonal splitting evolves with strain. Currently, calculations in Ref.~[\onlinecite{Bhandari2019}] suggest a lowering of the $J=3/2$-band at this point, while the measurements in Ref.~[\onlinecite{Paris2020}] indicate the opposite.

Works modelling greater tetragonal elongation in a superexchange model, such as  Ref.~[\onlinecite{Perkins2014}], have explored regimes our work did not. In those regimes the canting angle is supressed by the distortions. Ref.~[\onlinecite{Torchinsky2015}] found that the angles of the octahedral rotation and of the canting moment followed each other more closely with an additional staggered splitting between sublattices. Since we did not consider tetragonal splitting as a function of strain, the effects of an increased or staggered splitting is beyond the scope of this work.

Another interesting aspect expected to be affected by strain and an external field is the tendency to develop superconductivity. The mixing of $J$-states and the appearance of additional bands at the Fermim level might indicate that a $J=1/2$ $d$-wave superconducting state is less likely to develop. It is possible, however, that while the $d$-wave order parameter is less likely, another pairing function which involves multiple bands will become favorable. This is beyond the scope of the current manuscript and will be studied elsewhere.

\section{Acknowledgments}

The authors would like to thank Zi Yang Meng for useful discussions. We acknowledge financial support from NSERC, RQMP, FRQNT, a grant from Fondation Courtois, a Canada Research Chair, and an Alexander McFee Fellowship from McGill University. Computations were made on the supercomputers Beluga from \'Ecole de technologie sup\'erieure, and Cedar from Simon Fraser University, managed by Calcul Qu\'ebec and Compute Canada. The operation of these supercomputers is funded by the Canada Foundation for Innovation (CFI), the ministère de l'\'economie, de la science et de l'innovation du Qu\'ebec (MESI) and the Fonds de recherche du Qu\'ebec - Nature et technologies (FRQ-NT).

\appendix

\section{Self-Consistency Equations} \label{sec:appA}
In the mean field analysis, the order parameters are defined as the expectation values of bilinear operators calculated for the mean field eigenstates $|n (\boldk) \rangle$. Each order parameter is given by $\langle c^{\dagger}_{\gamma_{1}} c_{\gamma_{2}} \rangle_{s}$, where $\gamma_{i}$ is the label of one of the 6 local creation/annihilation operators given by $\alpha = yz, xz, xy$, and $\sigma=\uparrow, \downarrow$, for each of the sublattices $s=A,B$. The self-consistent solution for all possible order parameters is found iteratively and simultaneously by solving the set of coupled self-consistency equations:
\begin{equation}
\begin{array}{r l}
\langle c^{\dagger}_{\gamma_{1}} c_{\gamma_{2}} \rangle_{s} &= \disp \frac{1}{N} \dsum_{\boldk}^{N} \dsum_{n}^{12} \langle n (\boldk) | \gamma_{1}, s \rangle \langle \gamma_{2}, s | n (\boldk) \rangle n_{\text{F}}\left[ E_{n}(\boldk)\right] \\
&= \disp \frac{1}{N}\dsum_{\boldk}^{N} \dsum_{n}^{12} \eta^{\ast}_{\gamma_{1}, s, n} (\boldk) \eta_{\gamma_{2}, s, n} (\boldk) n_{\text{F}}\left[ E_{n}(\boldk)\right]
\end{array}
\end{equation}
where $n_{\text{F}}$ is the Fermi-Dirac distribution and the eigenvalues are given, for each $\boldk$ value, in the three-orbital basis, $|\gamma, s \rangle$, as $|n(\boldk) \rangle = \dsum_{\gamma, s} \eta_{\gamma,s, n} (\boldk) |\gamma, s \rangle$.

\section{Transformation into the $J$-basis} \label{sec:appB}
The order parameters are expressed in two alternative bases. The spin and orbital angular momenta are expressed in the basis of the three $t_{2g}$ orbitals. The other basis considered is the total angular momentum $J$-basis, which is the eigenstates in the large $\lambda$ limit. The transformation from the orbital and spin basis to the total angular momentum basis which is used in Eq.~\eqref{eqn:Utransf}, i.e., $\tilde{c}_{ m, \tau} = \sum_{\alpha, \sigma} U^{\alpha, \sigma}_{ m, \tau} c_{\alpha, \sigma}$, is given by 
\begin{equation} \label{eqn:Umat}
U = \left(
\begin{array}{cccccc}
 0 & \frac{1}{\sqrt{3}} & 0 & -\frac{i}{\sqrt{3}} & \frac{1}{\sqrt{3}} & 0 \\
 \frac{1}{\sqrt{3}} & 0 & \frac{i}{\sqrt{3}} & 0 & 0 & -\frac{1}{\sqrt{3}} \\
 0 & \frac{1}{\sqrt{6}} & 0 & -\frac{i}{\sqrt{6}} & -\sqrt{\frac{2}{3}} & 0 \\
 \frac{1}{\sqrt{6}} & 0 & \frac{i}{\sqrt{6}} & 0 & 0 & \sqrt{\frac{2}{3}} \\
 0 & \frac{1}{\sqrt{2}} & 0 & \frac{i}{\sqrt{2}} & 0 & 0 \\
 \frac{1}{\sqrt{2}} & 0 & -\frac{i}{\sqrt{2}} & 0 & 0 & 0 \\
\end{array} 
\right) 
\end{equation}
where $\boldsymbol{c}=( c_{yz, \uparrow}, c_{yz, \downarrow}, c_{xz, \uparrow}, c_{xz, \downarrow}, c_{xy, \uparrow} , c_{xy, \downarrow}) $ and $\tilde{\boldsymbol{c}} =( \tilde{c}_{1, +} , \tilde{c}_{1, -} , \tilde{c}_{2, +}, \tilde{c}_{2, -} , \tilde{c}_{3, +}, \tilde{c}_{3, -} )$. The new basis is $\tilde{c}_{m, \tau}$ where $m: 1=|1/2, \pm 1/2 \rangle, 2=|3/2, \pm 1/2 \rangle, 3=|3/2, \pm 3/2 \rangle$ and the pseudospin projections are labelled by $\tau=\pm$.

\begin{figure}
\centering 
\includegraphics[width=8.6cm]{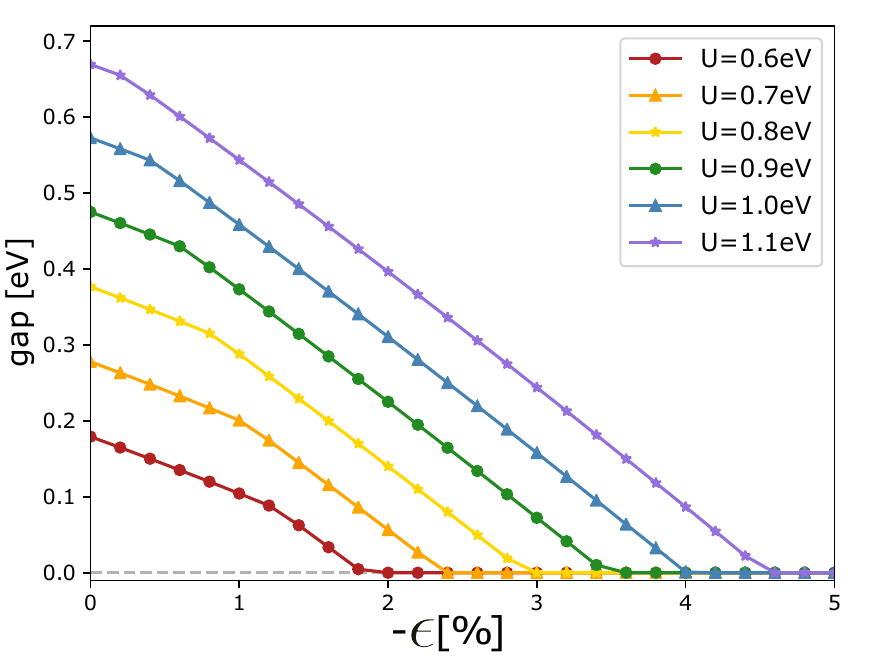} 
\caption{An increasing compressive strain, $\epsilon < 0$, decreases the initial insulating antiferromagnetic order. The critical strain, the value at which the gap closes, will be determined by the value of the gap at zero strain. The gap is plotted for different values of the interaction parameter $U$, with a Hund's coupling set to $J_{\text{H}}/U=0.1$. As the mean field approximation overestimates the order at zero strain, several vales of $U$ within the expected range are considered to get a possible range of values for the critical strain. Similarly as for the gap, the AFM order remains present at higher strains as $U$ is increased.}
\label{fig:gapStrain}
\end{figure}

\section{Critical strain values} \label{sec:appC}
The parameter choice of $U=0.9$eV, $J_{\text{H}}/U=0.1$, and $\lambda=0.38$eV, is used for the calculation in Fig.~\ref{fig:MomentumHx}. The values are close to the middle of the possible range for the Hund's coupling, $J_{\text{H}}/U=0.05-0.2$, and the spin-orbit coupling, $\lambda=0.3-0.7$eV, and has a value $U$, as well as chosen to have a gap at zero strain close to that found in experiments $\Delta_{c} = 0.35–0.65$eV\cite{Zhou2017, Watanabe2010,Wang2015, Kim2008, Kim2009, Kim2012}. The critical strains, the values at which the strain-driven phase transitions occur for compressive strain, are directly dependent on the size of the initial gap. The initial gap depends on the strength of the various interaction terms, the SOC $\lambda$, the Hund's coupling $J_{\text{H}}$, and the Zeeman field. Therefore the critical strain values increase with $\lambda$ and $U$, and decrease with $J_{\text{H}}$.

In Fig.~\ref{fig:gapStrain} we present results for calculations of the gap when the compressive strain is increased, for a range of possible values of the interaction $U$. The values for $U$ are those which have replicated the zero strain band structure using other methods. As a mean field analysis tends to overestimate the antiferromagnetic order we find a gap corresponding to experimental values at zero strain for a smaller $U$ than other methods do\cite{Nishiguchi2019, Meng2014, Lenz2019}. The experimental compressive strain values\cite{ Seo2019} reach up to $\epsilon=-1.9 \%$, so a quantitative prediction of the transition into a metallic state should be found at higher compressive strain values. Stronger interactions $U$ predict higher critical strain values while going through the same phase transitions. Within the limits of the mean field approximation, a prediction of a realistic band structure at zero strain and the value for critical strain will be a trade-off, and therefore a range of possible values are given here.

\begin{figure}
\centering 
\includegraphics[width=4.1cm]{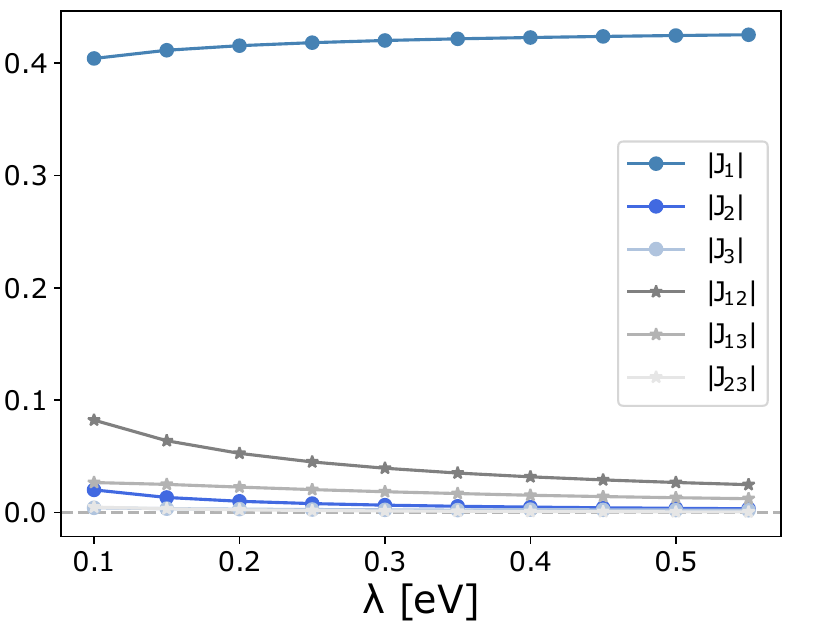} 
\includegraphics[width=4.1cm]{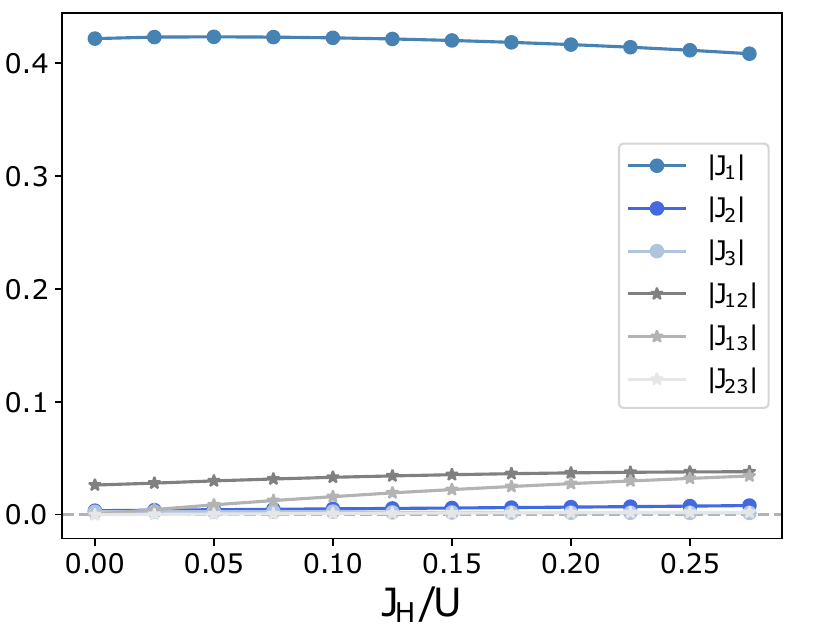} 
\caption{The three-orbital model used in this work allows us to consider how the contributions from each orbital changes for different sets of interaction strengths. The contributions to the staggered AFM order are shown at varying spin-orbit coupling $\lambda$ at $U=0.9$eV and $J_{\text{H}}/U=0.1$ as well as for varying Hund's coupling $J_{\text{H}}$ at $U=0.9$eV and $\lambda=0.38$. A higher $\lambda$ separates out the $J=1/2$ bands from the rest, resulting in a larger dominance of the $J_{1}$ contribution, as defined in Eq.~\eqref{eqn:Jm}. A larger Hund's coupling $J_{\text{H}}$ increases interorbital contributions and results in a larger mixing between $J$-sectors, as given in Eq.~\eqref{eqn:Jmn}.}
\label{fig:JnetlaJH}
\end{figure}

In Fig.~\ref{fig:JnetlaJH} the contributions to the staggered moment are considered, with no strain, for some additional values of the spin-orbit coupling $\lambda$ and the Hund's coupling $J_{\text{H}}$. For a higher SOC the $J=1/2$ states, $J_{1}$ as defined in Eq.~\eqref{eqn:Jm}, become clearly more dominant as the $J_{1}$ net moment increases in magnitude while the other contributions decrease. This is to be expected as the SOC separates the remaining bands from those of mainly $J=1/2$ character. A higher Hund's coupling the $J=1/2$ states instead become less dominant as the contribution remains constant while the mixing between $J$-states increases.

\begin{figure} 
\centering 
\includegraphics[width=8.6cm]{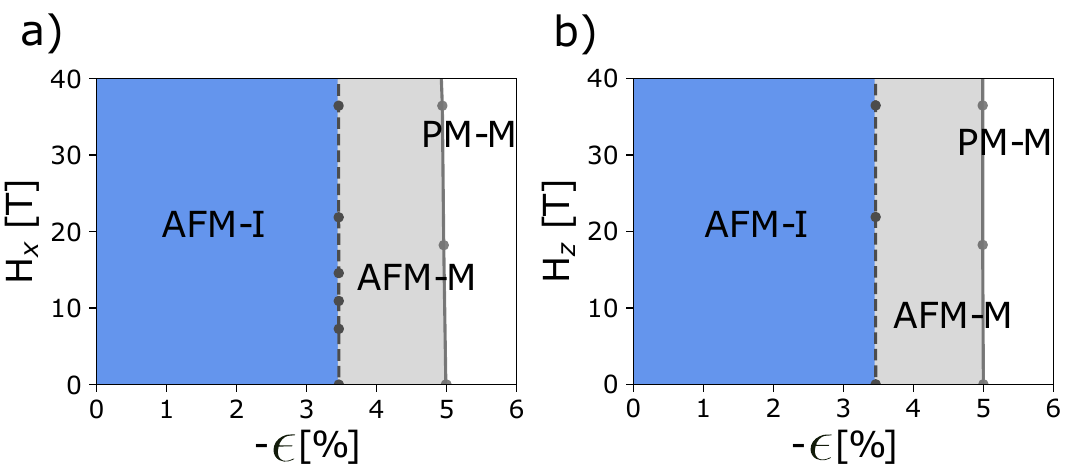} 
\caption{Phase transitions from the insulating AFM (AFM-I) state into metallic states occurs under compressive epitaxial strain $\epsilon < 0$. Phase diagrams are presented for a) an in-plane field along the $a$-direction $(H_{x})$, and b) the field is in the out-of-plane $z$-direction. As in Fig.~\ref{fig:MomentumHx} the AFM order decreases under an increasing strain until the indirect gap closes into a metallic order (AFM-M) and eventually goes through a first-order transition into a paramagnetic state (PM-M). The Zeeman field offers a minimal shift of the phase boundaries.}
\label{fig:PDsepH}
\end{figure}

\section{Phase diagrams with Zeeman field} \label{sec:appD}

A Zeeman field only has minor effects on the gap closing and the transition from the metallic AFM order to the paramagnetic state. The main effect of a Zeeman field on the strain-induced transitions is to lower the critical strain value, by reducing the indirect gap. The orbital and spin content of each band vary around some points of the Brillouin zone, which is shown in Fig.~\ref{fig:bandStrucs}. Therefore, a Zeeman field allows for the manipulation of the band structure with possible gap closures at various points in momentum space. An in-plane field ($H_{x}$) increases the band splitting around the $M$-point of the Brillouin zone and an out-of-plane field ($H_{z}$) results in an increased splitting at the $\Gamma$-point. In the phase diagrams in Fig.~\ref{fig:PDsepH}, where compressive strain and a Zeeman field has been applied, it is however apparent that even a large field can only modify the critical strain by an amount around $0.01\%$. The second transition, from the antiferromagnetic metallic (AFM-M) order into the paramagnetic metal (PM-M), occurs when the antiferromagnetic order parameters have reached a low enough value. An out-of-plane field results only in a small modification of the antiferromagnetic order and the second transition remains largely unchanged. An in-plane field has a slightly larger effect due to its effect on the canting angle and can shift the transition point further, yet still to a minimal amount. Although any shifts of transition points are difficult to achieve in Sr$_{2}$IrO$_{4}$, due to the large fields required, their effects might be of interest in other systems with similar characteristics.

\bibliography{MF_paper_refs}

\end{document}